\documentclass[twocolumn,aps,floatfix]{revtex4}
\usepackage{amssymb} \usepackage{graphicx} \usepackage{amsmath}
\usepackage[T1]{fontenc} \usepackage{pstricks} \usepackage{subfigure}
\usepackage[colorlinks=true,citecolor=blue,urlcolor=red]{hyperref}
\usepackage[normalem]{ulem}
\usepackage{hyperref}
\usepackage{doi}
\usepackage{comment}

\begin{document}
\title{Dynamics of large samples of repulsive Fermi gases at nonzero temperatures}


\author{Jaros{\l}aw Ryszkiewicz, Miros{\l}aw Brewczyk, and Tomasz Karpiuk}
\affiliation{\mbox{Wydzia{\l} Fizyki, Uniwersytet w Bia{\l}ymstoku,  ul. K. Cio{\l}kowskiego 1L, 15-245 Bia{\l}ystok, Poland } }

\date{\today}

\begin{abstract}

We develop a model of a binary fermionic mixture, consisting of large number of atoms, applicable at nonzero temperatures, in the normal phase. We use this approach to study dynamics of degenerate Fermi systems under various perturbations. For example, we analyze spin-dipole oscillations of a two-component fermionic mixture, demonstrating that the ferromagnetic phase shows up at stronger repulsion between components while the temperature raises. 
We study as well the radial oscillations of weakly interacting repulsive Fermi gases. We obtain a good agreement with experimental data when available. Otherwise, we compare our results with the outcome of the Hartree-Fock orbital calculations done for the system with small number of fermions.

\end{abstract}

\maketitle

\section{Introduction}

Systems of ultracold fermionic atoms have been already studied for years, both experimentally and theoretically. Since first experimental achievement of quantum degeneracy in fermionic potassium gas \cite{DeMarco99}, followed by successful attempts to cooling other elements \cite{Truscott01,Schreck01,Granade02,Hadzibabic03}, the interest in cold fermionic gases has quickly increased, covering broad range of quantum many-body phenomena including thermodynamic and transport related effects at unitarity \cite{Navon10,Ku12,Cao11}, correlations, in particular in optical lattices \cite{Rom06,Greif13,Kondov15,Schreiber15,Hart15,Cheuk16,Salomon19,Chiu19} or near Feshbach resonances \cite{Greiner05,Sanner12,Amico18}, strongly interacting gases in lower dimensions \cite{Vogt12,Koschorreck13,Luciuk17,Murthy19,Luick20,Bohlen20}, or dipolar gases \cite{Lu12,Aikawa14,Frisch14,Burdick16}.

In particular, dynamics of fermionic gases has been thoroughly investigated. The measurement of collective mode frequencies and damping rates as a function of temperature supplied evidences for superfluid behavior of a Fermi gas \cite{Kinast04}, while an observation of a vortex lattice \cite{Zwierlein05} provided direct verification for the superfluidity. By exciting hydrodynamic modes, such as collective oscillations \cite{Kinast04,Bartenstein04,Altmeyer07}, sound \cite{Joseph07}, or rotational modes \cite{Clancy07}, transport properties of a unitary Fermi gas have been experimentally determined. Careful analysis of compression, quadrupole, and scissors modes in the unitarity limit in the range of temperatures above the critical temperature for superfluidity has been performed in \cite{Riedl08} and revealed transition from hydrodynamic to collisionless behavior with increase of temperature. Recently, oscillations of repulsive binary fermionic mixture \cite{Sommer11,Valtolina17}, initially phase-separated by domain wall, were studied experimentally in connection with the long-standing problem of Stoner instability \cite{Stoner33}.

In this paper we investigate finite temperature dynamics of Fermi-Fermi mixtures by using the density-functional like description. To derive the equations of motion we start with introducing the semi-classical distribution function for fermions. Then we evoke the Kohn and Sham \cite{Kohn65} way of treating nonzero temperatures case within the density-functional methods and replace the local kinetic energy expression by the one corresponding to the free energy. Next we switch to the quantum hydrodynamic description \cite{Madelung27} of the system, apply the inverse Madelung transformation \cite{Dey98,Domps98,Grochowski17}, and follow Dirac prescription \cite{Dirac30} to get desired equations.

The paper is then organized as follows. First, we present the model of a two-component Fermi gas in the normal phase capable to retrieve dynamics when the number of atoms is large (Section \ref{distribution}). To prove the effectiveness of our model we compare numerical results to experimental data on dynamics of fermionic systems in the case of spin-dipole modes \cite{Valtolina17} (Section \ref{spindipole})
and to the outcome of the Hartree-Fock orbital calculations in the case of radial oscillations of weakly interacting repulsive Fermi gas (Section \ref{radialmodes}). We conclude in Section \ref{conclusion}.

\section{Equations of motion}
\label{distribution}

A simple description of a one-component gas in terms of a semi-classical distribution function $f_{\bf{p}}({\bf{r}})$ assumes that $f_{\bf{p}}({\bf{r}}) d{\bf{r}} d{\bf{p}}/(2\pi\hbar)^3$ gives the mean number of particles in the phase-space volume element $d{\bf{r}} d{\bf{p}}$. At equilibrium, at a given temperature $T$ and a chemical potential $\mu$, one has for a degenerate Fermi gas
\begin{eqnarray}
f_{\bf{p}}({\bf{r}}) = \frac{1}{e^{[\varepsilon_{\bf{p}}({\bf{r}})-\mu]/k_B T} + 1}
\label{distfunction}
\end{eqnarray}
with $\varepsilon_{\bf{p}}({\bf{r}})$ being the particle energy at position ${\bf{r}}$. For a single component ideal Fermi gas in a trap this energy becomes
\begin{eqnarray}
\varepsilon_{\bf{p}}({\bf{r}}) = \frac{{\bf{p}}^2}{2m} + V_{tr}({\bf{r}})  \,.
\label{spenergyideal}
\end{eqnarray}
The density of particles is obtained by integrating the distribution function over all momenta $n({\bf{r}}) \sim \int f_{\bf{p}}({\bf{r}}) d{\bf{p}}$.
Other kinds of energies can be added to Eq. (\ref{spenergyideal}), in particular the one related to the Weizs\"acker correction $E_W=\xi(T)\, (\hbar^2/2m) \int (\nabla \sqrt{n({\bf{r}})})^2\, d{\bf{r}}$, with a weakly temperature dependent coefficient $\xi(T)$ \cite{Perrot79}. Now, when the other fermionic component comes to the scene, the interaction energy has to be included as well. Assuming the inter-component interactions depend on densities only $V_{int}(n_+,n_-)$ (hereafter the components are distinguished by indices '$+$' and '$-$'), Eq. (\ref{spenergyideal}) becomes
\begin{eqnarray}
\varepsilon_{\bf{p}}({\bf{r}}) = \frac{{\bf{p}}^2}{2m} + V_{tr}({\bf{r}}) + \frac{\delta E_{W}^+}{\delta n_+} + \frac{\delta V_{int}}{\delta n_+}   \,.
\label{spenergyWei}
\end{eqnarray}

The density of particles of '$+$' component is calculated as
\begin{eqnarray}
n_{+}({\bf{r}}) &=& \int \frac{1}{e^{[\varepsilon_{\bf{p}}({\bf{r}})-\mu_+]/k_B T} + 1}\,  \frac{d{\bf{p}}}{(2\pi\hbar)^3}
\nonumber  \\
&=& \frac{1}{\lambda^3} f_{3/2}(z_+) 
\label{density}
\end{eqnarray}
and the energy density related to the local motion as
\begin{eqnarray}
\varepsilon_{+}({\bf{r}}) &=& \int \frac{{\bf{p}}^2/2m}{e^{[\varepsilon_{\bf{p}}({\bf{r}})-\mu_+]/k_B T} + 1}\,  \frac{d{\bf{p}}}{(2\pi\hbar)^3} \nonumber  \\
&=& \frac{3}{2} \frac{k_B T}{\lambda^3} f_{5/2}(z_+)    \,,
\label{energydensity}
\end{eqnarray}
where $k_B$ is the Boltzmann constant, $\lambda=\sqrt{2\pi\hbar^2/m k_B T}$ is the thermal wavelength, and $f_{3/2}(z)$ and $f_{5/2}(z)$ are the standard functions for fermions \cite{Huang}. The 'extended fugacity' equals
\begin{eqnarray}
z_+({\bf{r}}) = e^{(\mu_+ - V_{tr}({\bf{r}}) - \delta E_W^+/\delta n_+ - \delta V_{int}/\delta n_+ )/ k_B T}  \,.
\label{zet}
\end{eqnarray}
The chemical potential $\mu_+$ is determined by the normalization condition $N_+=\int n_{+}({\bf{r}}) d{\bf{r}}$.

According to the Kohn and Sham proposition \cite{Kohn65} on a generalization of the density-functional formalism to finite temperature case, for further analysis the energy of the system should be replaced by its free energy, whose density is
\begin{eqnarray}
f_{+}({\bf{r}}) = \frac{k_B T}{\lambda^3} \left[ (\ln{z_+})\, f_{3/2}(z_+) - f_{5/2}(z_+) \right]  \,.
\label{freeenergy}
\end{eqnarray}
Additionally, in a dynamical case the energy functional has to be modified by adding the energy of a macroscopic flow.
Then the part of the functional related to '$+$' component, which is minimized to get the equations underlying the system's dynamics, becomes
\begin{eqnarray}
F_+(n_+,{\bf{v}}_{+}) &=& \int f_{+}({\bf{r}})\, d{\bf{r}} + \int n_+ \frac{1}{2} m {\bf{v}}_{+}^2 d{\bf{r}} 
\nonumber  \\
&+& \int V_{tr}({\bf{r}})\, n_+\, d{\bf{r}} + E_W + V_{int}  \,.
\label{energyplus}
\end{eqnarray}
Here, the ${\bf{v}}_{+}({\bf{r}})$ is the velocity field of a macroscopic flow of '$+$' fermionic component and the second term on the right-hand side represents the energy of such motion.

Now we introduce the pseudo-wave function $\psi_+({\bf{r}})$ ($n_+=|\psi_+|^2$) for '$+$' component in such a way that
\begin{eqnarray}
\frac{\hbar^2}{2m} (\nabla \psi_{+}^*)\,  (\nabla \psi_{+}) = \frac{\hbar^2}{2m}  
(\nabla |\psi_{+}|)^2  +  n_{+} \frac{1}{2}\, m\, {\rm v}_{+}^2  \,.
\label{splitting}
\end{eqnarray}
The functional Eq. (\ref{energyplus}) is then transformed to
\begin{eqnarray}
&&F_+(\psi_+,\nabla \psi_+) = \int f_{+}({\bf{r}})\, d{\bf{r}} + \int \Big(-\frac{\hbar^2}{2m} \psi_+^* \nabla^2 \psi_+\Big)\, d{\bf{r}}  \nonumber  \\ 
&&- \frac{\hbar^2}{2m} \int (\nabla |\psi_+|)^2 d{\bf{r}} + \int V_{tr}({\bf{r}})\, n_+\, d{\bf{r}} + E_W + V_{int}  \,. \nonumber  \\ 
\label{energyplusbis}
\end{eqnarray}
Similar functional applies to the second component. The equations of motion are
\begin{eqnarray}
i \hbar\, \frac{\partial}{\partial t} \psi_{\pm}({\bf r},t) = \frac{\delta}{\delta \psi_{\pm}^*} 
F_\pm[\psi_{\pm}, \nabla \psi_{\pm}]  \,.
\label{EQmotion}
\end{eqnarray}
Since
\begin{eqnarray}
\frac{\delta}{\delta n_\pm} \int f_{\pm}({\bf{r}})\, d{\bf{r}} = k_B T\, \ln{z_\pm}   \,,
\label{functionalder}
\end{eqnarray}
the equations of motion become
\begin{eqnarray}
i \hbar\, \frac{\partial \psi_{\pm}}{\partial t} &=& 
\left(-\frac{\hbar^2}{2m} \nabla^2 + \frac{\hbar^2}{2m} \frac{\nabla^2 |\psi_{\pm}|}{|\psi_{\pm}|} +
k_B T\, \ln{z_\pm}  \right.  \nonumber  \\
&+& \left.\, V_{tr} - \xi(T)\, \frac{\hbar^2}{2m} \frac{\nabla^2 \sqrt{n_\pm}}{\sqrt{n_\pm}}  + \frac{\delta V_{int}}{\delta n_\pm}  \right) \psi_{\pm} \,. \nonumber  \\ 
\label{equmotion}
\end{eqnarray}
While solving Eqs. (\ref{equmotion}), the extended fugacities $z_\pm({\bf{r}})$ are found from the self-consistency condition $f_{3/2}(z_\pm) = \lambda^3\, n_\pm$, Eq. (\ref{density}), with $n_\pm=|\psi_\pm|^2$. The Weizs\"acker correction (the one before the last one) becomes less important when the number of atoms increases.

\section{Spin-dipole oscillations}
\label{spindipole}

We first examine our model in the case of experiment on spin-dipole oscillations of repulsive two-component fermionic mixtures \cite{Valtolina17}. This experiment was aimed to prove the existence of the phase transition from paramagnetic to ferromagnetic phase in a system of two-component short-range repulsive Fermi gas, i.e., to resolve the long-standing hypothesis of itinerant ferromagnetism posed by E. Stoner \cite{Stoner33}. It was predicted in \cite{Stoner33} that not localized electrons get into ferromagnetic state when short-range repulsion between opposite spin electrons becomes large enough to beat the Fermi pressure. 

To minimize the effect of pairing phenomenon \cite{Sanner12,Amico18}, in the experiment of \cite{Valtolina17} a mixture of ${^6}$Li atoms was prepared in a special state, in which both components were spatially separated. It was realized in two steps. First, components held in a prolate harmonic trap were spatially separated by using a magnetic field gradient. Next, when the overlap of two components was small enough, the optical repulsive barrier separating clouds was switched on and the magnetic field gradient turned off. Then the optical barrier was suddenly removed and the spin dynamics, i.e. oscillations of centers of mass of each component, was studied. Both frequencies and damping rates were measured, which demonstrated the existence of the critical repulsion between components. For weak repulsion the effect of softening of the spin-dipole mode was observed, i.e., the frequency of oscillations was continuously decreasing with a strength of a repulsion. Simultaneously, both atomic clouds were passing through each other. For stronger repulsion, however, qualitatively different behavior was found. Two atomic clouds started to bounce off each other with frequency higher than the axial trap frequency.

\begin{figure}[htb] 
\includegraphics[width=7.4cm]{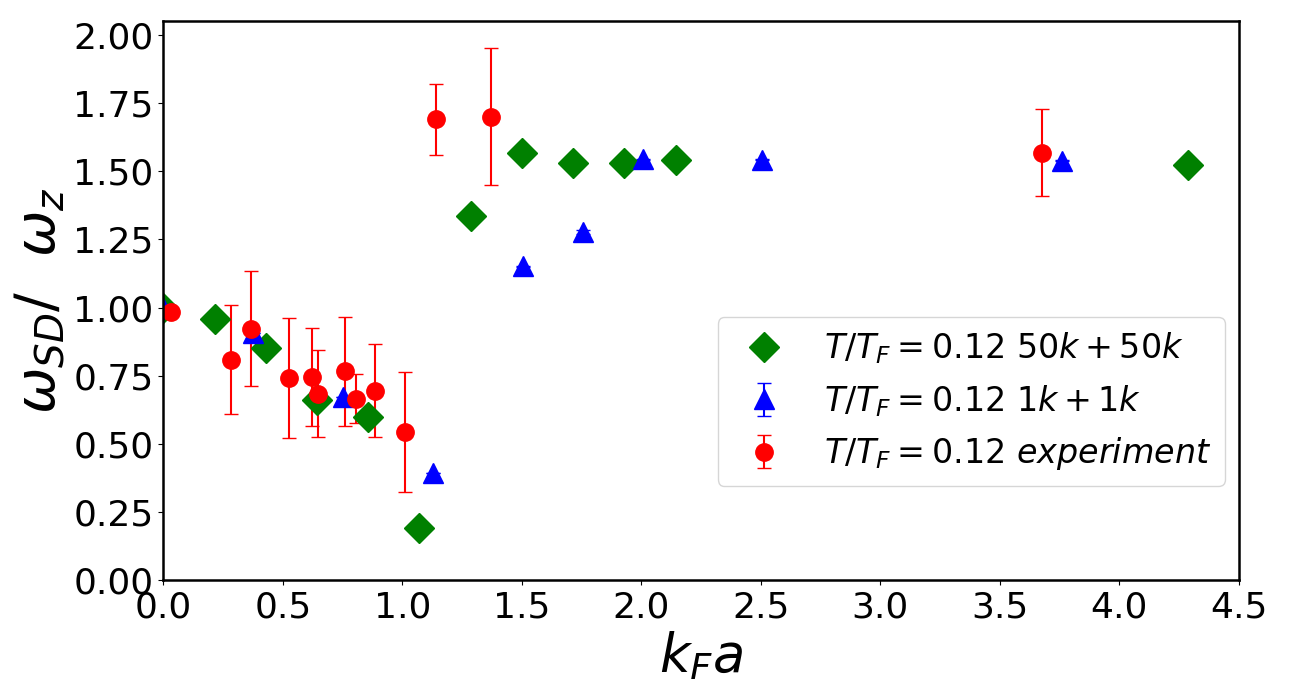} \\
\includegraphics[width=7.4cm]{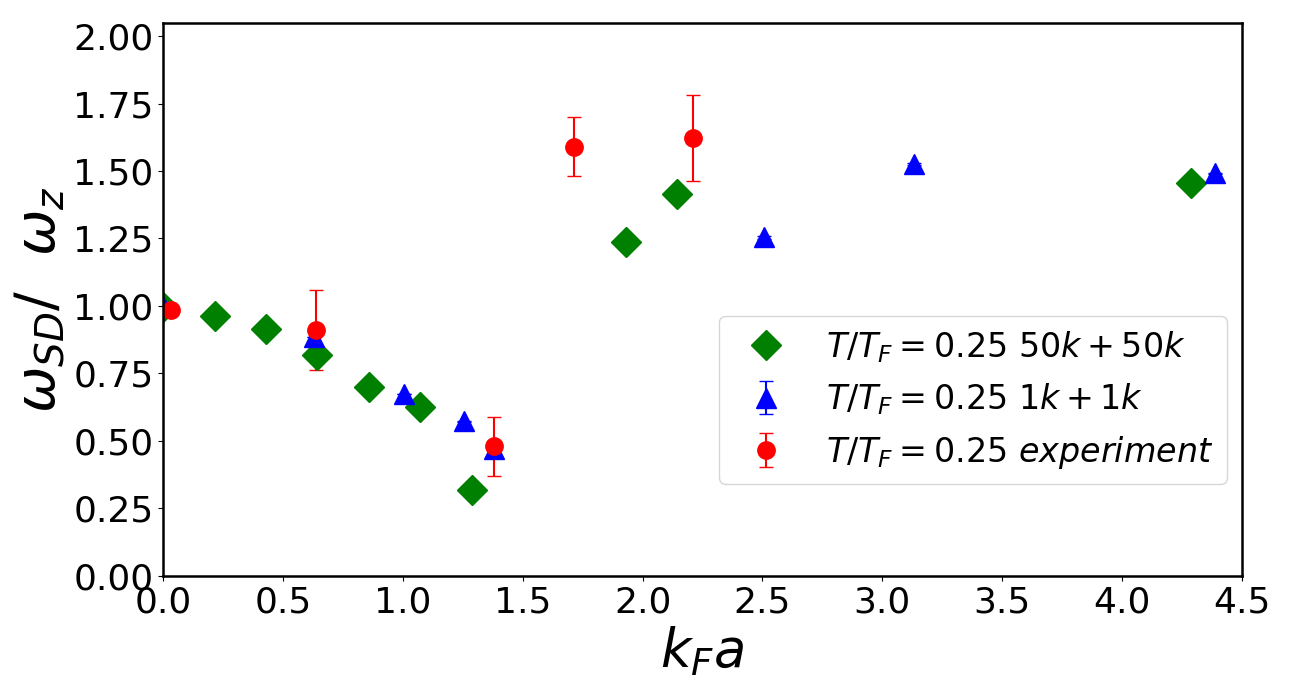}
\caption{Frequencies of the spin-dipole mode of a repulsive two-component Fermi gas plotted as a function of $k_F a$ for the temperatures $T/T_F=0.12$ (upper frame) and $T/T_F=0.25$ (lower frame) -- a comparison with the experiment of Ref. \cite{Valtolina17}. Simulations were performed for the system with number of atoms equal to $N/2=50000$ (as in the experiment), here the Weizs\"acker correction in Eq. (\ref{equmotion}) can be safely neglected, and $N/2=1000$. }  
\label{sd-hydr-exp}
\end{figure}

To model the experiment of Ref. \cite{Valtolina17} with Eqs. (\ref{equmotion}), we first obtain the initial state of two-component Fermi gas by solving Eqs. (\ref{equmotion}) by imaginary time technique \cite{Gawryluk18} at the presence of a cigar-shaped harmonic trap with radial and axial frequencies equal to $\omega_{\perp}=2\pi \times 265\,$Hz and $\omega_z=2\pi \times 21\,$Hz, respectively, and at the presence of the repulsive optical barrier. Then we remove the barrier and monitor the dynamics of the system by calculating the relative distance $d(t)$ between centers of two spin clouds. Analyzing $d(t)$ as a function of time we extract both the frequency and the damping rate of the mode (see Ref. \cite{Ryszkiewicz20} for details).

To get an agreement with experimental data we must include in the model the many-body correlations due to interactions. This can be achieved by renormalizing the coupling constant in the two-body contact potential \cite{Stecher07,Grochowski17,Ryszkiewicz20}. For a uniform system, it is done in a way to get correct low-density expansion, in parameter $\tilde{k}_F a$, of an energy of a two-component Fermi gas, see Eq. (3) in Ref. \cite{Ryszkiewicz20}. Here, $\tilde{k}_F$ is the Fermi wave number and $a$ is the $s$-wave scattering length. For a trapped gas a local density approximation is used. In the mean-field approximation the interaction energy density is $g n_+ n_-$, with $g=4\pi \hbar^2 a/m$, and the term $\delta V_{int}/\delta n_\pm=g n_\mp$ appears in Eqs. (\ref{equmotion}). After renormalization, $g n_\pm$ term is replaced by $g n_{\pm} + A (4/3\, n_{\mp}^{1/3}\, n_{\pm} + n_{\pm}^{4/3}) + B (5/3\, n_{\mp}^{2/3}\, n_{\pm} + n_{\pm}^{5/3})$ with $A=3 g a (6\pi^2)^{1/3} (11-2\ln{2}) /35 \pi$ and $B=3 g a^2 (6\pi^2)^{2/3} \pi/4 \times 0.23$ \cite{Grochowski17}.

In Fig. \ref{sd-hydr-exp} we show numerical results for frequencies of the spin-dipole mode for two temperatures studied experimentally in \cite{Valtolina17} (see Fig. 2). In the upper frame additional experimental point (most right) is included, see Fig. 3\,c in Ref.  \cite{Valtolina17}. Simulations were performed for the system both with the number of atoms as in the experiment ($50$ thousand of atoms in each component) and much smaller ($N/2=1000$). Fig. \ref{sd-hydr-exp} proves overall agreement between numerics and experimental data. Our calculations reveal softening of the spin-dipole mode followed by the transition from the paramagnetic to ferromagnetic phase. The hydrodynamic model, we developed, gives correct value of $k_F a$ (here, $k_F=(24 N)^{1/6}/(\hbar/m\, \overline{\omega})^{1/2}$, where $\overline{\omega}$ is the geometric mean of trap frequencies in all directions) at which this transition occurs. It is already well understood that at zero temperature the softening phenomenon depends solely on the combination $k_F a$ \cite{Recati11}. Our simulations support this observation also for nonzero temperatures, see  Fig. \ref{sd-hydr-exp}. According to Stoner's model the transition to ferromagnetic phase depends on $k_F a$ only as well, which again is exhibited by our simulations.

Above the critical value of $k_F a$ both components stop to penetrate each other and oscillate with frequencies smaller than twice the axial trap frequency, in agreement with experiment. Numerical results seem to be consistent for two considered numbers of atoms.
Our observation is, however, that the size of the intermediate regime (the one between the paramagnetic and ferromagnetic regimes) differs depending on the number of atoms in the sample. For larger systems the large-$(k_F a)$ value of the oscillation frequency is reached faster, i.e., for smaller $k_F a$. This can be understood as follows. For small samples ($N=1000$ in our case) even after crossing the critical value of $k_F a$ we can still observe the gas transmission through the intercomponent interface on the perimeter. Hence, in the intermediate regime the flow is partially still miscible. The full transition into the immiscible regime (i.e., when the oscillation frequency takes its large-$(k_F a)$ value) is then shifted to stronger interactions. This transition occurs faster (in terms of $k_F a$) for larger systems. It happens because, first, the damping rates for spin-dipole oscillations are high for the values of repulsion strengths $k_F a$ close to the critical one (see Fig. 3 in Ref. \cite{Ryszkiewicz20}), and, second, these rates are bigger for systems with larger number of atoms. Since damping rates decrease with temperature, the intermediate region broadens with temperature.


\begin{figure}[htb] 
\includegraphics[width=7.4cm]{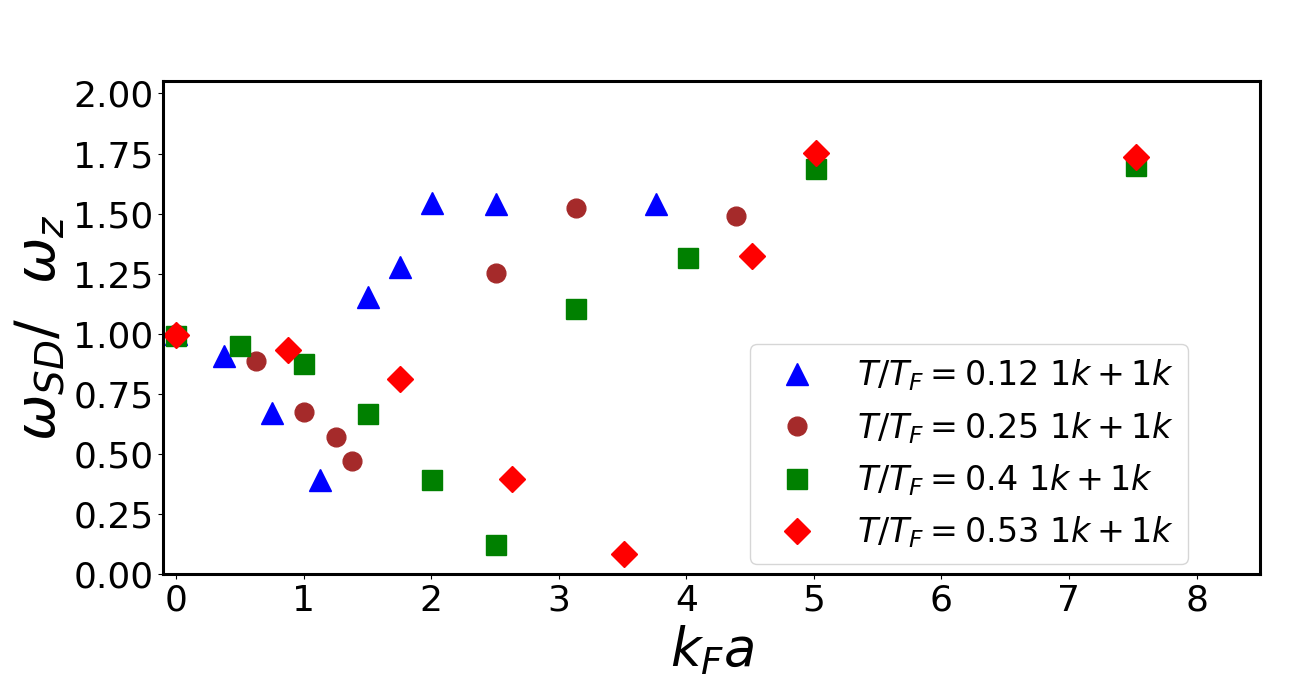}  \\ \vspace{0.4cm}
\includegraphics[width=7.cm]{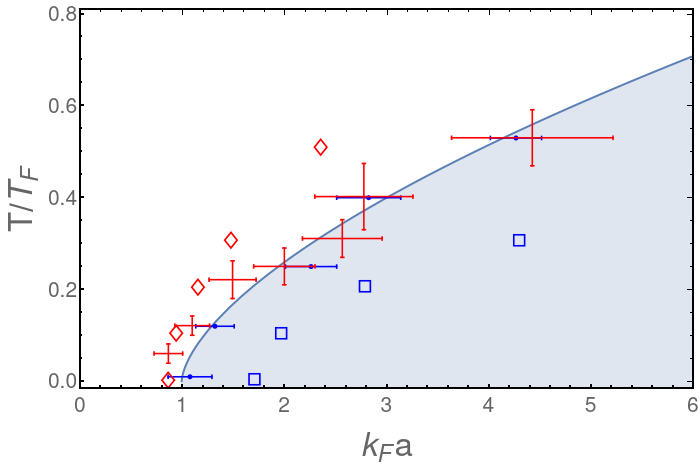}
\caption{Upper frame: Frequencies of the spin-dipole mode of a repulsive two-component Fermi gas plotted as a function of $k_F a$ for different temperatures, up to $T/T_F=0.53$. Lower frame: Phase diagram showing the critical value of the repulsive interaction strength at a given temperature. The red crosses are the experimental data taken from Ref. \cite{Valtolina17} (Fig. 3 d), while blue bullets come from numerics. Solid line, which is a power-law fit to the numerical points, separates the paramagnetic (white area) from the ferromagnetic (dark area) phase. Blue squares and red diamonds are the predictions of static Stoner's model assuming interactions between atoms are not and are renormalized, respectively. }  
\label{diagram}
\end{figure}

Fig. \ref{diagram} summarizes hydrodynamic calculations performed for smaller atomic samples (Fig. \ref{sd-hydr-exp} demonstrates that the transition to ferromagnetic phase occurs actually at the same value of $k_F a$, independently of the number of atoms). In the upper frame we show the frequencies of the spin-dipole mode for additional two temperatures, $T/T_F=0.4$ and $T/T_F=0.53$, for the system consisting of $1000$ atoms in each component. These temperatures were studied in  \cite{Valtolina17}, although in a different way -- not by following the spin-dipole oscillations but by analyzing the stability of initially created spin domains. The lower frame is the phase diagram, it gathers the results corresponding to transition between the paramagnetic and ferromagnetic phases. This diagram shows the critical value of $k_F a$ as a function of temperature. The experimental data are marked as red crosses, taken from Ref. \cite{Valtolina17} (Fig. 3\,d), while numerical results are put as blue bullets. The solid line is a power-law fit to the numerical points and separates the paramagnetic and ferromagnetic phases. At a given value of $k_F a$ the ferromagnetic phase is entered while the system's temperature is decreased, in qualitative agreement with Stoner's model of itinerant ferromagnetism -- the ferromagnetic phase rises when interactions are able to overcome the fermionic quantum pressure, which gets lower with decreasing temperature.

To check on a quantitative level our numerical (dynamic) results versus predictions of original (static) Stoner's model we compare the kinetic energy of the gas to its interaction energy at equilibrium \cite{Zwerger09}. In the simplest case, i.e., in the mean-field approximation the interaction energy is $g \int n_+ n_- \, d{\bf r}$ and the above mentioned comparison reads
\begin{eqnarray}
\frac{3}{2} \frac{k_B T}{\lambda^3} \int f_{5/2}(z_+)\, d{\bf r} =  \frac{4 \pi \hbar^2}{m\, k_F}  \left( \int n_+ n_- \, d{\bf r} \right)\, (k_F a)_{cr}  \,.  \nonumber  \\
\label{comparison}
\end{eqnarray}
The critical value of repulsive interactions $(k_F a)_{cr}$ is found assuming equal component densities $n_+ = n_- = f_{3/2}(z_+)/\lambda^3$ normalized to $N_+=N_-=1000$ with $z_+=\exp{[(\mu_+ - V_{tr})/ k_B T]}$. The critical values $(k_F a)_{cr}$ as a function of $T/T_F$ are plotted in Fig. \ref{diagram}, lower frame, as blue squares. At zero temperature $(k_F a)_{cr} \approx 1.7$ in agreement with our earlier calculations \cite{Grochowski17}, see Fig. 1d. When interactions between atoms get renormalized, the condition (\ref{comparison}) changes into the third degree polynomial equation for the critical interactions $(k_F a)_{cr}$. The solutions as a function of $T/T_F$ are shown in Fig. \ref{diagram}, lower frame, as red diamonds. Now, at zero temperature $(k_F a)_{cr} \approx 0.9$, in agreement with experimental data of Ref. \cite{Valtolina17} at the lowest temperature. Overall behavior of $(k_F a)_{cr}$ resembles the one determined experimentally and obtained in numerical simulations (both representing the dynamical Stoner effect), especially for lower temperatures.

A note regarding consistency of presented results with those reported already in Refs. \cite{Grochowski17,Ryszkiewicz20} is now in order. First, the frequency of the spin-dipole mode in a ferromagnetic phase strongly depends on the geometry of the trapping potential, while a density functional method is used. In elongated trap, as in the experiment of Ref. \cite{Valtolina17}, it is about $1.7$ (see Fig. \ref{diagram}, upper frame). At zero temperature and in a spherically symmetric trap this frequency approaches the value of twice the trap frequency \cite{Grochowski17}, with recognizable admixture of other frequency $(\omega_{SD}/\omega_z = \sqrt{2})$. On the other hand, within the Hartree-Fock approach the spin-dipole mode frequency in a ferromagnetic phase remains twice the axial frequency, independently of temperature \cite{Ryszkiewicz20}. This probably happens because our treatment of Hartree-Fock dynamics at nonzero temperatures does not allow for atoms to change between single-particle orbitals during the evolution. In our case only one-particle orbitals change in time during dynamics, not the populations -- populations are chosen by using a Monte Carlo sampling technique, before the barrier separating components is removed (for an approach in which populations are treated on the same way as orbitals, although at equilibrium only, see \cite{Lipparini}).

\section{Oscillations of weakly interacting repulsive fermionic mixtures}
\label{radialmodes}

In this section we carry out simulations of dynamics of two-component weakly interacting Fermi gas, initially confined in a spherically symmetric trap, after a weak disturbance of the trapping potential. Within a weak-driving regime the system's response can be treated analytically in the range of high temperatures and in the limit of an ideal gas.
Here, we are verifying our description of large Fermi systems by studying the monopole oscillations of a Fermi gas for weak repulsive interactions. As in Section \ref{spindipole}, we use renormalized interaction to describe two-component weakly interacting Fermi gas.

\begin{figure}[t!hb] 
\includegraphics[width=7.5cm]{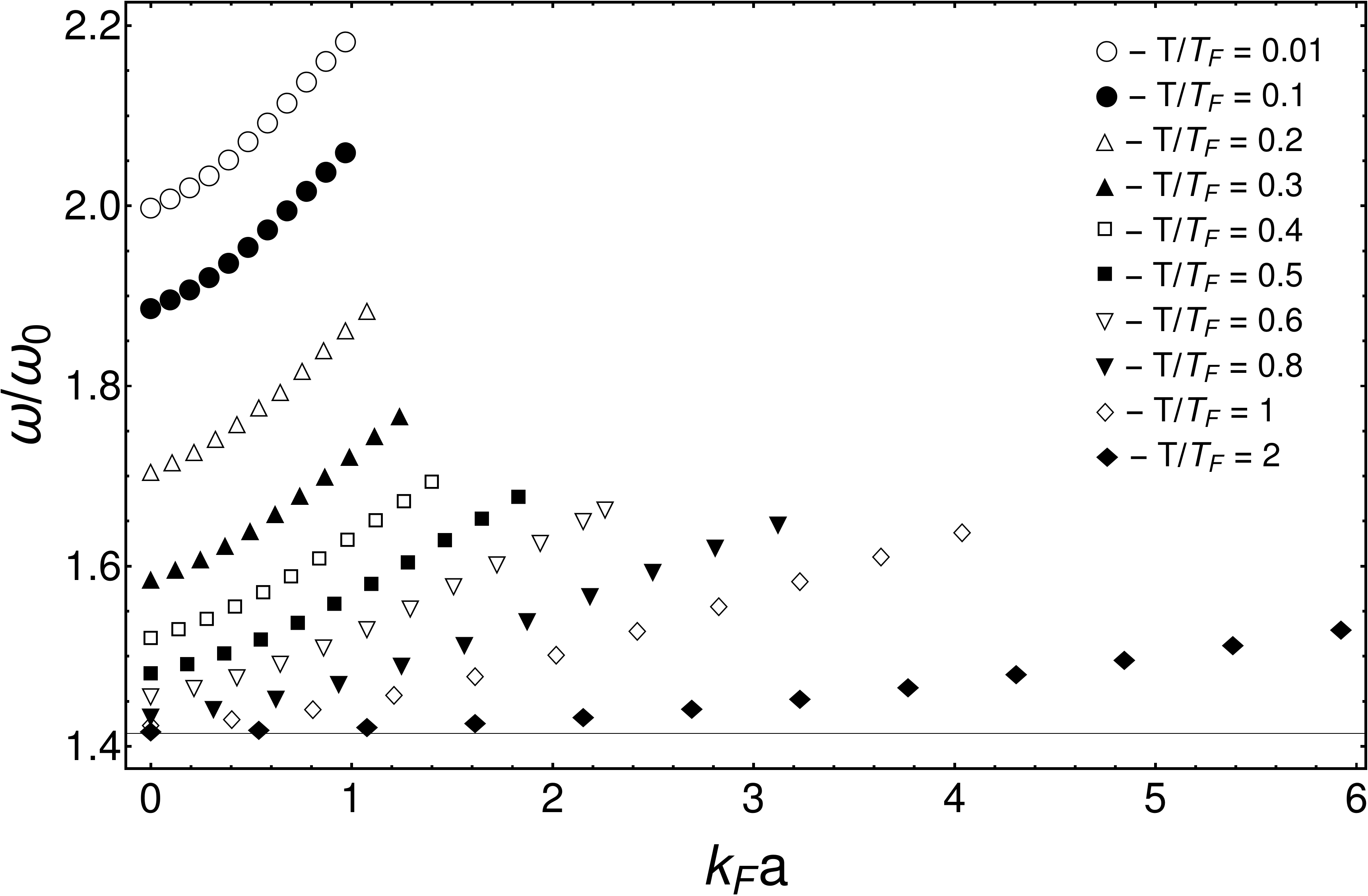}  \\ \vspace{0.4cm}
\includegraphics[width=7.5cm]{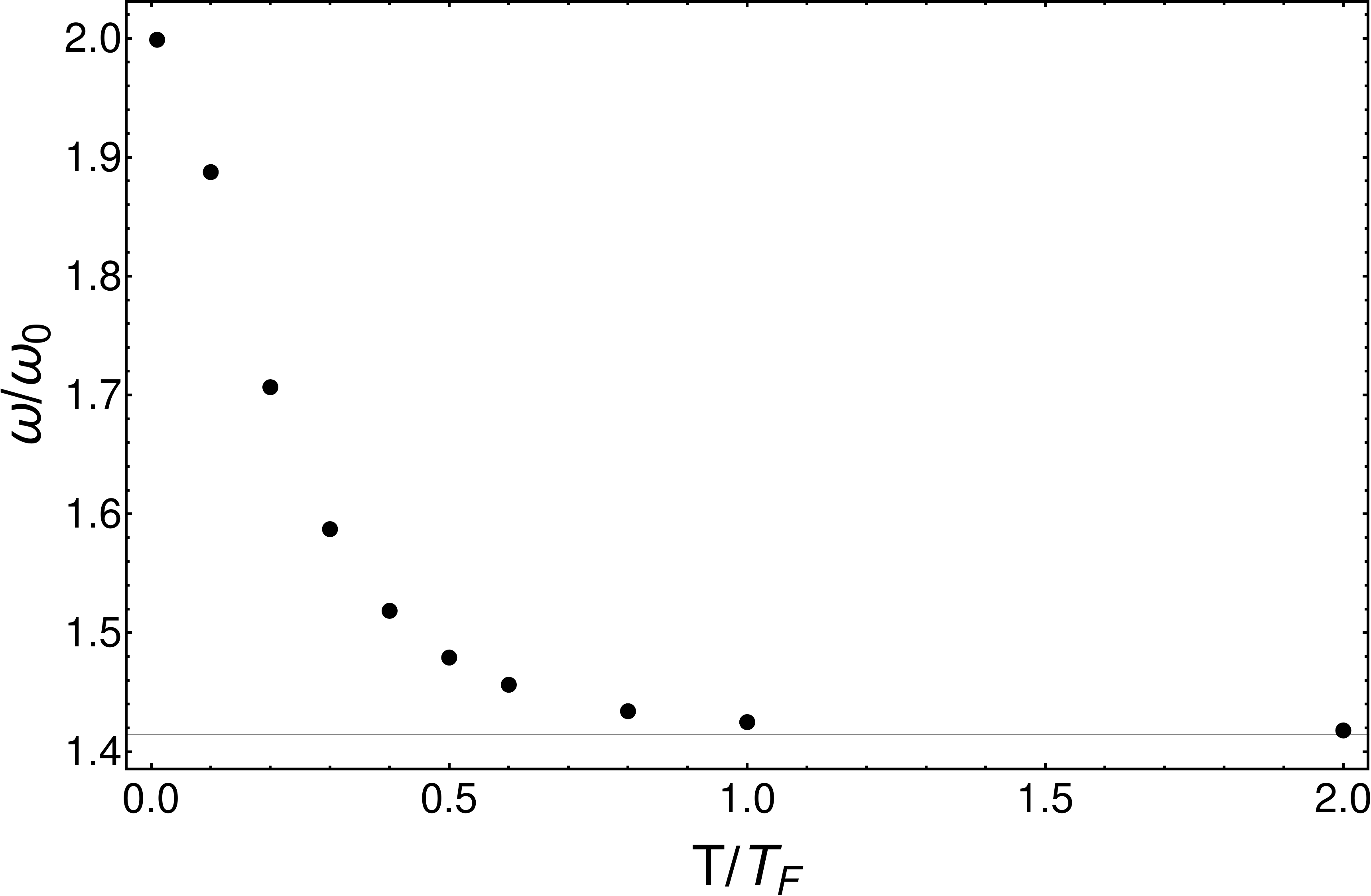}
\caption{Upper frame: Frequencies of in-phase monopole modes as a function of temperature. The critical interaction for which the phase separation occurs depends on temperature. Only frequencies of modes excited in the paramagnetic phase are shown. Both components consist of $1000$ atoms. Lower frame: The monopole mode frequency as a function of temperature in the limit of no interaction between components, clearly approaching $\sqrt{2}\, \omega_0$ for high temperatures. }  
\label{monopole}
\end{figure}

Both gases are perturbed in-phase, i.e., two atomic clouds are first simultaneously being squeezed as an effect of increasing trap frequencies and next the trapping potential is slightly attenuated (by decreasing trap frequencies) to allow the gas to expand. Such a cycle is repeated a few times after which the system starts to oscillate in a trap. In this way the spherically symmetric oscillations are excited. We find frequencies of such excitations by calculating the width of an atomic cloud $\int d^3{\bf r}\, r^2 n_{\pm}({\bf r},t)$ and analyzing its time dependence. We show frequencies of in-phase monopole modes as a function of temperature in Fig. \ref{monopole}, limiting ourselves only to the paramagnetic range of parameters \cite{Karpiuk20,Trappe16,Trappe21}. At zero temperature (open circles data) our results perfectly match those obtained within time-dependent Hartree-Fock method (see Ref. \cite{Karpiuk20}). In the limit of no interaction between components, the spherically symmetric mode oscillates with frequency $2\omega_0$, where $\omega_0$ is the trap frequency. While moving towards the paramagnetic-ferromagnetic phase crossing this frequency increases to about $2.2\omega_0$. For higher temperatures all mode frequencies are shifted down and in the limit of $T \gtrsim T_F$ can be studied analytically.

To analyze small in-phase oscillations of two-component interacting Fermi gas at high temperature limit we utilize the Madelung representation \cite{Madelung27} of Eqs. (\ref{equmotion}). For the '$+$' component this representation is expressed as a set of equations for the density and the velocity fields:
\begin{eqnarray}
&&\frac{\partial n_{+}}{\partial t} + \nabla \cdot (n_+ {\bf v}_+) = 0 \nonumber  \\
&&m \frac{\partial {\bf v}_+}{\partial t} + \nabla \Big( k_B T\, \ln{z_+} + V_{tr} + \frac{\delta V_{int}}{\delta n_+} + \frac{1}{2} m {\bf v}_{+}^2 \Big) = 0 \,. \nonumber  \\
\label{Madelung}
\end{eqnarray}
At high temperature limit one has $n_+ \lambda^3=f_{3/2}(z_+) \approx z_+$. 
Small oscillations are investigated by assuming small deviations of the state from the equilibrium and by looking for periodic solutions for deviations. We write the density as $n_+=n_{eq}^+ + \delta n_+$, where $\delta n_+$ is the departure from the equilibrium density and assume that both the velocity and $\delta n_+$ are small quantities. Since $n_{eq}^+= \exp{[(\mu_+ - V_{tr} - (\delta V_{int}/\delta n_+)_{eq})/k_B T]}/\lambda^3$ and $\delta n_+=\delta n_-$ (in-phase oscillations), the Eqs. (\ref{Madelung}) are transformed to
\begin{eqnarray}
\frac{\partial}{\partial t} \delta n_+ &=& - \nabla \cdot (n_{eq}^+\, {\bf v}_+)  \nonumber  \\
m \frac{\partial {\bf v}_+}{\partial t} &=& - \nabla \left[ k_B T\, \frac{\delta n_+}{n_{eq}^+} + G(n_{eq}^+)\, \delta n_+ \right]  \,,
\label{Madelung2}
\end{eqnarray}
where
\begin{eqnarray}
G (n_{eq}^+) = \left[ \left( \frac{\partial}{\partial n_+} + \frac{\partial}{\partial n_-} \right) \frac{\delta V_{int}}{\delta n_+} \right]_{n_+=n_-=n_{eq}^+}  \,,
\label{G}
\end{eqnarray}
and can be combined into a single equation for the density deviation
\begin{eqnarray}
m \frac{\partial^2}{\partial t^2} \delta n_+ = \nabla \cdot \left[ n_{eq}^+\,  \nabla
\left( k_B T\, \frac{\delta n_+}{n_{eq}^+} + G(n_{eq}^+)\, \delta n_+ \right) \right] \,.  \nonumber  \\
\label{Madelung3}
\end{eqnarray}

Now, the limit of small interactions can be analyzed. For weak inter-species interactions the second term in the right-hand side of Eq. (\ref{Madelung3}) is neglected and $n_{eq}^+= \exp{[(\mu_+ - V_{tr})/k_B T]}/\lambda^3$. Eq. (\ref{Madelung3}) can be rewritten as
\begin{eqnarray}
m \frac{\partial^2}{\partial t^2} \left( \frac{\delta n_+}{n_{eq}^+} \right) = k_B T\, \nabla^2 \left( \frac{\delta n_+}{n_{eq}^+} \right) - (\nabla V_{tr}) \nabla \left( \frac{\delta n_+}{n_{eq}^+} \right)   \,. \nonumber  \\
\label{Madelung4}
\end{eqnarray}
The trapping potential is spherically symmetric, $V_{tr}=m\, \omega_0^2\, r^2 /2$, and we search for periodic solutions $\delta n_+ / n_{eq}^+ \sim e^{-i \omega t}$ of Eq. (\ref{Madelung4}) which are spherically symmetric as well. The solutions can be found by using the power series method. The lowest energy mode has a frequency $\omega=\sqrt{2}\, \omega_0$, marked by a horizontal solid line in Fig. \ref{monopole}, lower frame. The mode itself is $\delta n_+ \sim (1-m\omega^2 r^2/(6 k_B T))\, n_{eq}^+ $.

\section{Conclusions}
\label{conclusion}

In summary, we have studied dynamics of mixtures of repulsive Fermi gases consisting of large number of atoms at nonzero temperatures. We find a quantitative agreement with experimental results of \cite{Valtolina17} on spin-dipole oscillations. The calculations show the dependence of the critical repulsion $k_F a$ on the temperature. The transition to the ferromagnetic phase requires larger value of $k_F a$ with increasing temperature, in agreement with Stoner's picture of itinerant ferromagnetism. We also model breathing modes of weakly interacting repulsive fermionic mixtures, getting agreement with low temperature results of \cite{Karpiuk20} and showing decrease of oscillation frequencies with increase of temperature.

\acknowledgments  
The authors acknowledge support from the (Polish) National Science Center Grant No. 2018/29/B/ST2/01308. Part of the results were obtained using computers at the Computer Center of University of Bialystok.

\end{document}